\documentclass[12pt]{article}
\usepackage{amsmath}
\usepackage{amssymb}
\usepackage{amsthm}
\usepackage{amscd}
\usepackage{amsfonts}
\usepackage{graphicx}%
\usepackage{fancyhdr}
\usepackage{color}
\usepackage[font={small}]{caption}

\theoremstyle{plain} \numberwithin{equation}{section}

\theoremstyle{definition}

\usepackage[left=1.5cm,bottom=2cm]{geometry}

\begin{document}

\noindent \begin{Large}{\bf Impact of triplet correlations on neural population codes} \\   \end{Large} \\
\noindent {\bf Short Title:} Impact of triplet correlations on neural population codes\\  \\
N. Alex Cayco-Gajic, Joel Zylberberg, Eric Shea-Brown\\
Department of Applied Mathematics, University of Washington\\

\section{Abstract}

Which statistical features of spiking activity matter for how stimuli are encoded in neural populations?  A vast body of work has explored how firing rates in individual cells and correlations in the spikes of cell pairs impact coding.  But little is known about how higher-order correlations, which describe simultaneous firing in triplets and larger ensembles of cells, impact encoded stimulus information.  Here, we take a first step toward closing this gap. We vary triplet correlations in small ($\approx 10$ cell) neural populations while keeping single cell and pairwise statistics fixed at typically reported values. For each value of triplet correlations, we estimate the performance of the neural population on a two-stimulus discrimination task. We identify a predominant way that such triplet correlations can strongly enhance coding:  if triplet correlations differ for the two stimuli, they skew the response distributions of the two stimuli apart from each other, separating them and making them easier to distinguish. This coding benefit does not occur when both stimuli elicit similar triplet correlations. These results indicate that higher-order correlations could have a strong effect on population coding. Finally, we calculate how many samples are necessary to accurately measure spiking correlations of this type, providing an estimate of the necessary recording times in experiments.

\section{Author Summary}
Traditional theories of neural coding rely on tuning curves describing the average response of a neuron to a stimulus.  Adding complexity to this neuron-by-neuron view is the fact that the spikes emitted by different cells in response to a given stimulus are often correlated. This covariability is often assessed across pairs of neurons and can have diverse and potentially strong impacts on how neural circuits encode stimuli.  

Recent experiments suggest that ``beyond-pairwise" spike correlations occur among larger groups of cells.  However, we have little understanding of how such correlations affect neural coding.  Here, for small ($\approx 10$ cell) populations, we isolate the effect of triplet correlations while keeping the spike rates and pairwise correlations fixed at typical reported values.  In a simple setting, we identify cases where the resulting triplet correlations do (and do not) have a significant effect on coding performance.  We explain our findings geometrically via the {\it skew} that triplet correlations induce in population-wide distributions of neural responses.  Finally, a major challenge of measuring correlated spiking is the large amount of data that is required for accurate detection. We estimate the length of recordings necessary to assess such effects empirically.



\section{Introduction}

The brain transforms sensory inputs into spiking activity that is distributed across neural populations and is variable from trial to trial.  What are the key statistical features of this activity that determine the amount of sensory information encoded by such a population?  Much can be learned by quantifying the mean responses as well as the trial-to-trial variability of spikes emitted by individual cells.  However, this variability is often coordinated across the population. Significant correlations between the spikes emitted simultaneously by pairs of cells have been observed across the brain, e.g. in visual cortex \cite{hansen, martin, kohn} (but see \cite{ecker}), auditory cortex \cite{decharms}, motor cortex \cite{maynard}, prefrontal cortex \cite{constantinidis}, the lateral geniculate nucleus \cite{alonso}, and retina \cite{mastronarde} -- possibly reflecting circuit mechanisms such as recurrent connectivity and common input \cite{trong,kohn,Bin+01,reid,shadlen,bruno}.  Such pairwise spike correlations can have a wide range of impacts on stimulus encoding.  In principle, pairwise correlations can interfere with population-wide averaging that would otherwise damp noise; conversely, they may play a more positive role, allowing variability to be cancelled or even acting as an extra conduit of information independent of firing rates.  Thus, a large body of theoretical work has been dedicated to understanding the precise relationship between pairwise correlations and population coding (e.g., \cite{zohary,abbott,averbeck,panzeri,oram,sompolinsky,dasilveira,hu,shamir14}).

Intriguingly, recent experiments suggest that knowing the correlations between pairs of neurons is not enough to characterize collective activity across a neural population. This implies the existence of ``higher-order correlations" (HOCs): that is, correlated firing between groups of three or more cells that is either more or less than what would be expected from the firing rates and pairwise correlations alone \cite{ganmor,ohiorhenuan,montani,shimazaki,koester,tkacik}.  Results to date illustrate that, as for pairwise correlations, HOCs can have a range of positive to negative effects on stimulus encoding.  This is assessed by comparing coding fidelity based on the ``full" responses recorded simultaneously in a population, with coding fidelity based on a model population that has the same firing rates and pairwise correlations but no HOCs.  In \cite{ganmor}, HOCs among retinal ganglion cells improved coding efficiency -- specifically, they increased the speed with which the identity of two types of visual stimulus could be decoded from the population response.  Meanwhile, in \cite{montani}, HOCs in somatosensory cortex decreased mutual information between neural activity in rat somatosensory cortex and the frequency of whisker stimulation.

These findings raise two important questions. First, when should we expect HOCs to have a significant impact on population coding? Second, are there simple rules of thumb that predict when they will facilitate versus hinder the population code? These questions remain largely unexplored, but the answers may lead to new perspectives on neural coding, as many studies to date have used measures of coding accuracy (such as the optimal linear estimator \cite{salinas}) that do not incorporate the effects of HOCs.  

Developing a general answer to these questions is difficult.  
First, note that there is an exponential growth in the number all possible higher-order correlations with the population size $N$.  Here, we limit our investigation to the first higher-order interaction beyond pairwise correlations -- that is, to triplet correlations. We make this choice for simplicity, but note that experimentally-observed HOCs in \cite{ganmor,montani} are limited to interactions among small groups of 3-5 cells, the most numerous of which are triplets. Additionally, 
we focus on relatively small populations, a point we return to in the discussion. Finally, we analyze two-stimulus discrimination tasks: specifically, we ask how a ``preferred" stimulus can be distinguished from a nearby ``nonpreferred" stimulus that elicits lower firing rates across a population.  

Throughout, we use maximum-entropy statistical models \cite{schneidman} that isolate the effect of triplet correlations, while fixing the lower-order statistics (i.e., mean activity of each neuron and correlations among each pair) to prescribed values typical of those reported in physiology experiments. By triplet correlations, we mean the probability of simultaneous spiking in triplets of cells beyond what is expected by the lower-order statistics. We first study populations with homogenous statistics, and then move to the heterogeneous case; for concreteness, we choose statistics drawn from the distributions observed in mammalian primary visual cortex \cite{martin}. We find that triplet correlations that are consistent with these constraints can strongly improve stimulus encoding, if they have a stimulus-dependent structure.  Specifically, if triplet correlations are between cells with similar stimulus tuning, and if they are larger for the nonpreferred versus the preferred case (or, to a lesser extent, vice-versa), then the triplet correlations will separate the distributions of spikes produced by each stimulus.  As a result, the stimuli can be better discriminated; in other words, the population carries more stimulus information. Comparable statistical models with stimulus-independent triplet correlations show relatively little effect on coding. We show that our results can be understood intuitively as either positively or negatively skewing the distribution of the summed population activity in short time windows. Our results show the importance of quantifying higher-order correlations in neurophysiology experiments, as they may have a significant impact on the function of neural systems. Finally, we give a simple calculation that estimates the length of recordings necessary to identify such triplet correlations experimentally.

\section{Materials and methods}

We investigate the effect of higher-order spike correlations (HOCs) on the level of stimulus information that a neural population encodes about pairs of stimuli:  a preferred stimulus (eliciting a higher firing rate), and a non-preferred stimulus.  Each stimulus elicits a different distribution of spike patterns characterized by firing rates, pairwise correlations, and HOCs. We vary the triplet statistics separately for each stimulus, and calculate the amount of information that spiking patterns contain about the stimulus identity. In order to isolate the effect of HOCs, we keep the lower-order statistics (i.e., firing rates and pairwise correlations) fixed during this process.  We do this by using a popular class of statistical models called maximum entropy models, which are able to match any given statistics of a population of neurons while minimally constraining other features of the spike distribution.

\subsection*{The maximum entropy model}

Consider the spikes emitted by $N$ cells in response to stimulus $S^{(m)}$, where $m=1$ or $2$.  Binning these spikes in small windows yields a sequence of spiking patterns $\vec\sigma$, each of which is a vector of 1s and 0s representing whether a given neuron spiked or not within that time window. Assuming that the population is at a stationary state, each pattern $\vec\sigma$ can be viewed as a random sample from a probability distribution that describes the simultaneous, population-wide response of the neurons to a particular stimulus.  These are the probability distributions that we will study in this paper.

If the $i^{th}$ neuron spikes with probability $\mu_i$ in each time window (i.e., the firing rate of the $i^{th}$ neuron is $\mu_i/\Delta t$), then the (simultaneous) pairwise spike correlations for cell $i$ and $j$ are:
\begin{equation}
\rho_{ij} = \frac{Pr( \sigma_i, \sigma_j=1) - \mu_i \mu_j}{\sqrt{\text{var}(\sigma_i) \text{var}(\sigma_j)}}.
\end{equation}
In other words, to quantify the correlation between pairs of neurons, one must subtract from the observed probability of simultaneous paired spiking the probability of simultaneous paired spiking in a ``null" model (in this case, assuming all activity is independent).

Similarly, quantifying higher-order correlations requires comparing against a null model. In this case, we use the pairwise maximum entropy model, which matches the observed lower-order statistics while making the fewest additional assumptions about the structure of the data \cite{schneidman03,schneidman}. Under this model, the probability of firing pattern $\vec\sigma$ under stimulus $S^{(m)}$ is given by:
\begin{equation} P_{PW}(\vec \sigma | S^{(m)}) = \frac{1}{Z} \exp\left[ \sum_i h^{(m)}_i \sigma_i +  \sum_{i>j} J^{(m)}_{ij} \sigma_i \sigma_j \right]. 
\label{e.pw}
\end{equation}
Here, the interaction terms $h_i^{(m)}$ and $J_{ij}^{(m)}$ are tuned so that the distribution matches the prescribed lower-order statistics, that is, firing rates and pairwise correlations. $Z$ is a normalization factor. Thus equipped, we define a measure of triplet correlations as the probability of three neurons firing simultaneously, relative to what would be expected from the pairwise maximum entropy model:
\begin{equation}
\kappa_{ijk} = Pr(\sigma_i, \sigma_j, \sigma_k = 1) - Pr_{PW}(\sigma_i, \sigma_j, \sigma_k = 1).
\label{e.kappa}
\end{equation}
We refer to this quantity as the ``excess triplet probability."

In order to explore the effects of HOCs, we add a triplet interaction term $G^{(m)}$ to the previous distribution:
\begin{equation} P(\vec \sigma | S^{(m)}) = \frac{1}{Z} \exp\left[ \sum_i h^{(m)}_i \sigma_i +  \sum_{i>j} J^{(m)}_{ij} \sigma_i \sigma_j +  G^{(m)} \sum_{i>j>k} \sigma_i \sigma_j \sigma_k \right]. 
\label{e.triplet}
\end{equation}
Increasing (or decreasing) $G^{(m)}$ increases (or decreases) the excess triplet probability $\kappa_{ijk}$.  For simplicity, we set the triplet interaction term to be the same for all triplets of neurons; however we have also added heterogeneity by adding zero-mean noise to the triplets $G^{(m)}$ terms for each triplet $i,j,k$, and we found the same qualitative results that we will report here, as long as the $G^{(m)}_{ijk}$ have the same sign for each triplet (data not shown).


The approach we have described is useful, because it allows us to isolate the effects of triplet correlations:  for each triplet interaction $G^{(m)}$, we re-fit the single-cell and pairwise interactions $h_i^{(m)}$ and $J_{ij}^{(m)}$ to maintain the same firing rates and pairwise correlations. However, this is computationally demanding, and limits the size of the populations that we can study systematically to around $N=10$ neurons.  We return to the issue of population size in the Discussion. 

\subsection*{Fitting the maximum entropy models}
To fit maximum entropy models \cite{jaynes}, we used improved iterative scaling (IIS), an algorithm that maximizes the average log-likelihood of the parameterized model to find the interaction parameters such that the moments of the resulting distribution match prescribed values \cite{berger,darroch}. For homogeneous populations, the interaction parameters $h_i^{(m)}$ and $J_{ij}^{(m)}$ are identical for each neuron and neuron pair.  Fitting is thus sped up considerably, as we are reduced to a three-parameter search. To explore the full range of possible triplet statistics that are consistent with prescribed single-cell and pairwise statistics, we varied the probability of synchronous triplet firing in steps of 0.001 and found the values for which the lower-order statistics and the probability of triplet firing converged within an average relative error of 1\% in 1000 steps. 
For heterogenous populations, we implemented a slight variant of this algorithm. We fixed the triplet interaction terms $G^{(m)}$, and then used IIS to tune the first and second order interaction terms so that the lower-order statistics converged to the specified values within an average 5\% error.


\subsection*{Mutual information between stimuli and firing patterns}
To quantify encoded stimulus information, we compute the mutual information between the binary firing pattern $\vec\sigma$ and stimulus $S^{(m)}$. This is given by the following difference in entropies:
\begin{equation}
I = H(\vec\sigma) - H(\vec\sigma|S).
\end{equation}
The first term denotes the entropy in the full distribution of firing patterns:
\begin{equation}
H(\vec\sigma) = - \sum_{\vec \sigma} P(\vec \sigma) \log_2 P(\vec \sigma).
\end{equation}
The second term, sometimes called the noise entropy, is the average entropy of the firing patterns conditioned on a particular stimulus (each of which we assume is equally likely):
\begin{equation}
H(\vec\sigma|S) = - \sum_{m} \sum_{ \vec \sigma} \frac{1}{2} P(\vec \sigma | S^{(m)}) \log_2 P(\vec \sigma | S^{(m)}).
\end{equation}
Thus, the mutual information quantifies how much entropy (or uncertainty) in the firing patterns is reduced given knowledge of the stimulus identity. The benefit of using mutual information is that it is not specific to a particular neural decoder. Instead, it can be thought of as an upper bound for how much information any decoder can extract from the spiking activity of the population. Throughout this paper, we calculate mutual information exactly, without requiring any entropy estimators.

To quantify the effect of beyond-pairwise statistics, we first calculate $I_{PW}$, the mutual information between the stimulus and the firing patterns of the pairwise maximum entropy models. This we compare to the information in populations that include triplet statistics with the following equation:
\begin{equation}
\text{relative }\Delta I = \frac{I-I_{PW}}{I_{PW}}.
\label{e.deltaI}
\end{equation}
This quantifies the factor of increase in mutual information that is gained by populations that include triplet statistics.

\subsection*{Homogenous populations}
As described above, we prescribe the firing rates and pairwise correlations in our neural populations and hold these statistics fixed while we vary triplet correlations.  We first consider populations with homogenous statistics: i.e., all neurons have the same firing rates, all pairwise correlations are the same, etc.  We consider various choices for the firing rates of our cells, in the range of 0.1 to 0.35 spikes per bin, with step sizes of 0.05.  For spikes counted in 20 ms bins, this corresponds to spiking at 5-17 Hz, a range similar to that of average stimulus-evoked firing rates under different preparations in rodent sensory cortex \cite{barth}.  We denote the difference in firing rates between the preferred and non-preferred stimulus by $\Delta \mu$, and use  values of $\Delta \mu =\mu^{(2)}-\mu^{(1)}= 0.05, 0.10, 0.15$ (2.5-7.5 Hz); larger values gave highly discriminable responses regardless of the choice of higher-order correlations.  We take pairwise noise correlations fixed at various values between 0 and 0.25, a range corresponding to values typically reported in, e.g., sensory and motor cortex \cite{cohen}.  For simplicity, we use the same values of pairwise correlations for both stimuli.

%


\subsection*{Heterogenous populations}

For populations with heterogenous spiking statistics, we make the following choices.  For concreteness, we choose firing rates and pairwise correlations  from distributions reported in anesthetized cat visual cortex in response to natural movies \cite{martin}.  Under the non-preferred stimulus, firing rates were taken to be exponentially distributed (as shown in \cite{baddeley}) with a median firing rate of 5 Hz as indicated in \cite{martin}. The activity under the preferred stimulus was given by adding to each cell's firing rate a Gaussian random variable with mean $\Delta \mu$ and standard deviation 0.02, where $\Delta \mu$ ranged from 0.1 to 0.15. 
The probability of spiking (or of two neurons spiking together) was constrained to be no less than 0.05 (2.5 Hz) to avoid convergence problems with tuning the maximum entropy models. 

Spike correlations between pairs of cells were drawn from a Gaussian distribution with mean and interquartile length of approximately 0.05 each, as reported for 20 ms time bins in \cite{martin}. These values were used as the elements in the spike count covariance matrix as long as they formed a positive semidefinite matrix; if the matrix were not positive semidefinite, another random draw of values was taken. Since larger correlations have been observed in other areas and preparations \cite{cohen}, we also repeated this study with average noise correlations of 0.1 and 0.2 and the same variance as before.  For simplicity, in all cases we continue to use the same noise correlation matrix for both stimuli.

All calculations were averaged over 24 random populations, i.e., 24 random draws from the same distributions of lower-order statistics.


\subsection*{Calculation of $T_\text{est}$}
Here we calculate the length of recordings that would be required in order to estimate a key quantity in our study:  the frequency with which three neurons fire within the same time bin. In particular, based on a particular experiment lasting $T$ time bins, we want to bound the 95\% confidence intervals of the relative error of the sample estimate of the frequency of cells $i,j,k$ firing within the same time bin in the data. Suppose we want the relative error between the estimated frequency $\hat p$ and the true frequency $p=Pr(\sigma_i,\sigma_j,\sigma_k=1)$ to be at most $\alpha$, which means the raw error must be bounded by $\alpha p$. Assuming the time bins are independent, the variance of the estimated frequency is $\text{var}(\hat p) = p(1-p)$. Under a normal approximation, the 95\% confidence interval for the true probability $p$ is within two standard errors above or below $\hat p$. This means that, in order to bound the relative error $(p-\hat p)/p$ by $\alpha$ with 95\% confidence, we must set the following inequality:
\begin{equation}
2 \sigma_\text{SEM} \leq \alpha p,
\end{equation}
Using the definition of the the standard error as $\sigma_\text{SEM} = \sqrt{\text{var}(\hat p)/T} = \sqrt{p(1-p)/T}$, this can be rearranged into the following equation for the desired length of the experiment:
\begin{equation}
T \geq \frac{1-p}{p\left(\frac{\alpha}{2}\right)^2}.
\label{e.test}
\end{equation}
The inequality above provides a lower bound on how many time bins are needed to estimate any triplet spike of probability of $p$ or greater within a relative accuracy of $\alpha$. In the text we call this lower bound $T_\text{est}$. 

\section{Results}
\begin{figure}[!t]
\begin{centering}
\leavevmode
\includegraphics[width=6.6in] {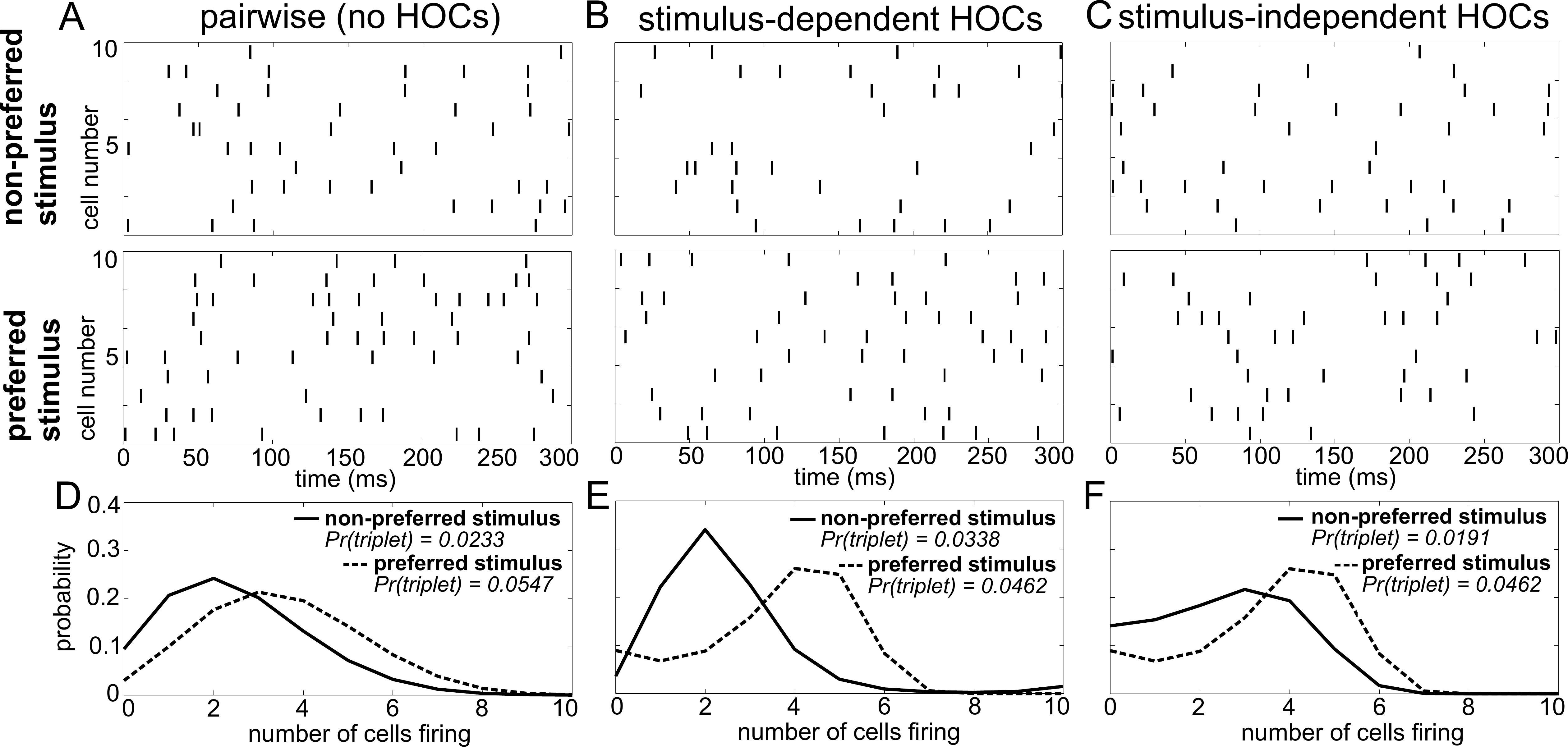}
\caption{Population spike responses in three examples with different higher-order correlations. (A-C) Raster plots for three sample populations in response to two stimuli (parameters are indicated in Figure 3A). All three populations have identical firing rates and pairwise correlations, and differ solely in the level of higher-order correlations.  (A) The ``pairwise" model, which can be fully described by the firing rates and pairwise correlations. In (B), the probability of three neurons spiking simultaneously has been increased (decreased) compared to the pairwise model in response to the non-preferred (preferred) stimulus. In (C), the probability of such triplet spiking is decreased for both stimuli. (D-F) Histograms of population spike count within 20 ms time bins for the three populations. Note how triplet correlations impact the skew of these response distributions (see text).}
\vspace{ 0 in}
\hspace{-.4 in}
\end{centering}
\end{figure}

Firing rates of individual neurons, and correlations between spiking activity in pairs of neurons, are the properties that are typically used in assessing neural variability and population coding.  Far less is known about the role of higher-order correlations (HOCs).  When and how should we expect HOCs to affect the fidelity of the neural code? 

As an example, Figure 1 shows spike trains of three sample populations in response to two different stimuli:  a preferred stimulus, eliciting relatively high firing rates, and a nonpreferred stimulus.   Importantly, all three of these populations have the same firing rates and pairwise correlations for each stimulus (i.e., the same ``lower order statistics"). The sole difference is in the HOCs within each population. In Figure 1A, the first and second order statistics are sufficient to fully characterize the responses.  That is, the responses follow a pairwise maximum entropy distribution~\cite{jaynes,schneidman03}. We refer to this simply as the ``pairwise" model; it is the null case against which we compare the responses of populations with other HOCs.  In Figure 1B, we modified the probability of three neurons firing within a short time window, keeping the lower-order statistics fixed. In particular, we changed triplet correlations in a stimulus-dependent way, so that the frequency of synchronous triplets is decreased under the preferred stimulus and increased under the non-preferred one. 

\begin{figure}[!t]
\begin{centering}
\leavevmode
\includegraphics[width=4in] {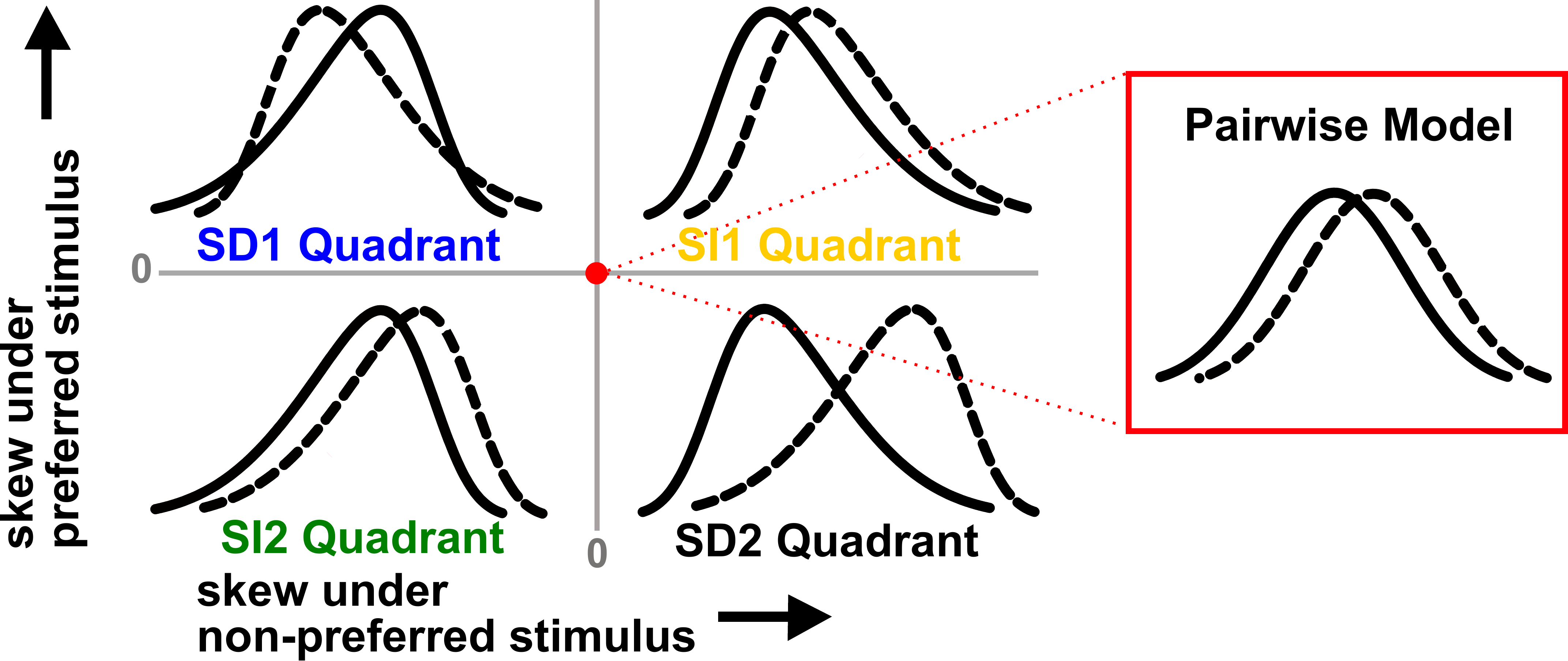}
\caption{Schematic illustrating how triplet correlations skew population spike count distributions.  Each quadrant corresponds to a different case of stimulus-dependent (SD) or stimulus-independent (SI) triplet spike correlations.  The means and variances of the distributions are the same for all four quadrants; only the skew differs (and higher moments). In particular, note that the distributions are pulled away from each other when the non-preferred response (solid line) is positively skewed and the preferred stimulus (dashed line) is negatively skewed (i.e., the SD2 quadrant).  This case gives the largest coding advantage (see text).}
\vspace{ 0 in}
\hspace{-.4 in}
\end{centering}
\end{figure}

It is difficult to visualize the difference in population spiking from the raster plots alone (e.g. comparing Figures 1A and 1B).  However, the implications for stimulus coding become apparent from distributions of the spike count, that is, the number of cells spiking within short time windows. For the pairwise model, these response distributions overlap strongly (Figure 1D).  Changing the triplet correlations significantly reduces this overlap by skewing the spike count histograms away from each other (Figure 1E).  Note that the stimulus dependence of the triplet correlations is crucial; simply changing the triplet correlations identically under each stimulus skews the spike count histograms in the same direction, preserving much of the overlap in the pairwise distributions (Figure 1CF).  This is the key observation from this example: increasing (or decreasing) the frequency of triplets of neurons firing together corresponds to increasing (decreasing) the skew of the spike count distribution, which can shape the response distributions to significantly improve stimulus encoding.  Moreover, the largest improvements arise when  triplet correlations for the two stimuli are distinct.

These observations are illustrated by the schematic in Figure 2. The labeled regions show the four possible types of skewed distributions for the preferred and non-preferred stimulus. If the signs of the triplet correlations are the same under each stimulus, we say they are stimulus-independent (SI). The skews of the spike count distributions then can either be larger compared to the pairwise model (which we call the SI1 quadrant), or smaller (SI2). Alternatively, the triplet correlations may be stimulus-dependent (SD), in which case they have opposite sign for the two stimuli (SD1 and SD2). Figure 2 shows that stimulus-dependent triplet correlations give a greater coding benefit than stimulus-independent ones. Moreover, the greatest benefit occurs in the SD2 quadrant, where the skewed distributions are the most strongly separated;  if neural populations produce responses of this type, ignoring HOCs may lead to a significant underestimation of encoded information. 

Guided by this intuition, we studied the range of effects that triplet correlations can have on encoded information in populations of $N=10$ neurons.  We first considered populations with homogenous firing rates and correlations for all cells, and then moved to the heterogeneous case, where we took lower-order statistics consistent with those observed in anesthetized cat V1 \cite{martin}.  In each case, we used maximum entropy models to manipulate the triplet correlations while keeping the lower-order moments fixed (see Methods).

\subsection*{Populations with homogenous statistics}

We first investigated populations with homogenous firing statistics (i.e., equal firing rates $\mu_i^{(m)} = \mu^{(m)}$, pairwise correlations $\rho_{ij}=\rho$, etc.). This simple case illustrates how the information in neural populations can vary with triplet firing statistics, and is used as a basis for studying more realistic populations in the next section.  As described above, we fixed the firing rates and pairwise correlations elicited by each stimulus, and independently varied triplet spike probabilities over the entire range for which the models can be tuned (see Methods for details). For each value of triplet correlation, we calculated the mutual information between the stimuli and the spike responses in the population.  Because the population is homogenous, this process simplifies:  a histogram of the total number of spikes produced in response to a stimulus (the spike count histogram) gives a complete representation of the population activity.  For example, the firing patterns 1010000100 and 0011001000 are equally likely to occur because they have the same number of active neurons.

\begin{figure}[!t]
\begin{centering}
\leavevmode
\includegraphics[width=6.6in] {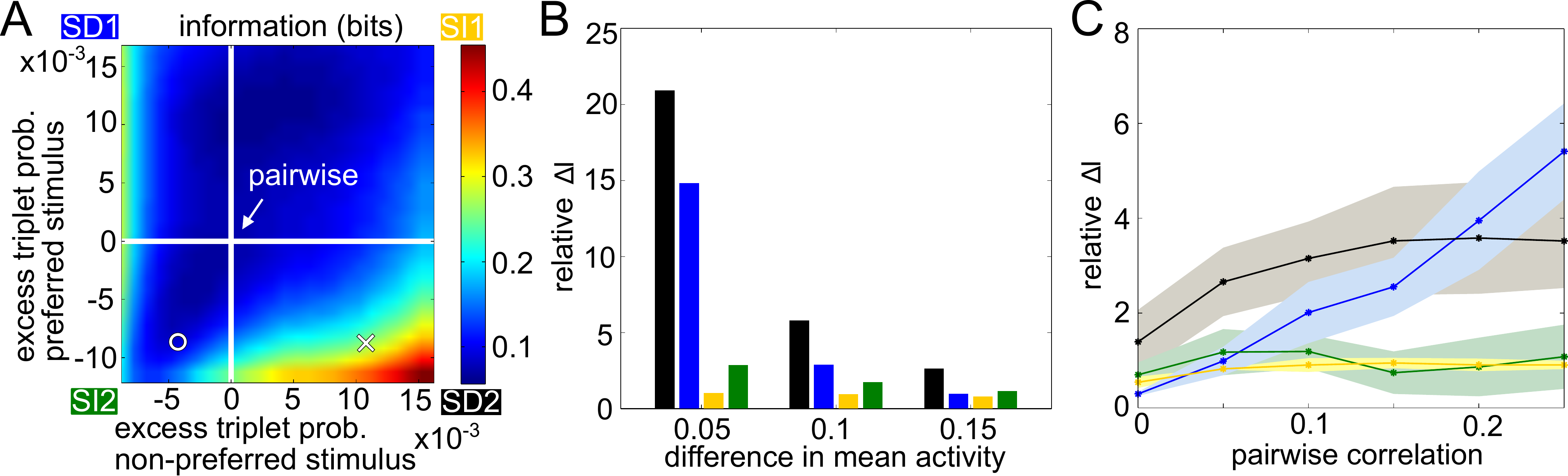}
\caption{Populations with homogenous statistics. (A) Mutual information in bits as the excess triplet probability is varied for responses to the preferred and non-preferred stimuli. White lines indicate the pairwise maximum entropy model under each stimulus (shown in Figure 1A). The cross marker indicates the population in Figure 1B; circular marker for Figure 1C. Quadrants are labeled corresponding to the different stimulus-dependent triplet correlations (see Figure 2). In this example, the firing rate $\mu_1 = 0.25$ for the non-preferred stimulus, $\mu_2 = 0.35$ for the preferred stimulus, and the pairwise correlation $\rho = 0.05$ for both stimuli. (B) Relative increase in mutual information for the full model compared to the pairwise fit (see text), averaged over populations with firing rates between $0.1 - 0.35$ but keeping $\Delta \mu$ fixed to 0.05, 0.10, or 0.15. Pairwise correlations are fixed to $\rho = 0.05$. Colors correspond to the corners of the quadrants indicated in A (blue, SD1; yellow, SI1, etc.). (C) Relative increase in mutual information as a function of pairwise noise correlations, averaged over different firing rates. Shading represents standard deviation over single-cell activity, ranging from $0.1 - 0.35$ with step sizes of 0.05.}
\vspace{ 0 in}
\hspace{-.4 in}
\end{centering}
\end{figure}

Figure 3A summarizes how triplet correlations can affect the level of encoded information in a homogenous population.  The axes of this plot are given by $\kappa$ (Equation~\eqref{e.kappa}), the excess probability of a triplet spike versus that expected in the corresponding pairwise model; they differ in scale because the range of realizable triplet spiking probabilities varies depending on the prescribed lower-order statistics.  Within this plot, the cross indicates the population illustrated in Figure 1BE, while the circle marker represents that in Figure 1CF. The pairwise distributions occur along the white lines; at their intersection is the case shown in Figure 1AD. The asymmetry between quadrants SD1 and SD2 is due to the difference in the average firing rate evoked by each stimulus.

The overall trends in mutual information agree with the intuition developed in Figure 2. Mutual information is largely increased with the presence of oppositely signed triplet correlations that skew the response distributions away from each other, whereas simply increasing or decreasing the triplet correlations independent of stimulus identity does not have a significant effect. This is especially true in the SD2 quadrant. In general, the relative effects on mutual information are strongest when the population activity is noisy relative to the difference in firing rates, i.e., when firing rates are similar under the two stimuli or when the correlation between pairs of cells is large (Figure 3BC). 

One concern is that our results for $N=10$ neurons may not hold for larger populations. To test this, we repeated our calculations of mutual information with fixed lower-order and triplet statistics, for increasing population size (up to $N=40$; see Supplementary Figure S1). We found that, for fixed $\kappa$, the relative increase in information can be stable across a range of population sizes, at least for homogenous populations; in fact, it increases slightly with $N$.  We return to the question of population size in the discussion.

\subsection*{Populations with heterogeneous statistics}
To test the effect of triplet correlations on stimulus encoding in a more realistic setting, we next considered populations with heterogenous statistics. For concreteness, we chose distributions of firing rates and pairwise correlations that have been observed in mammalian V1 (see Methods, Heterogeneous Populations). The difference in the average firing rates under each stimulus is a free parameter that determines the baseline level of encoded information in the pairwise models. If the stimulus-evoked firing rates are very different, higher-order correlations would have little room to improve discrimination. We therefore set $\Delta \mu$ so that stimulus discrimination was 60\% accurate on average for the pairwise models; later in this section this parameter was increased to correspond to up to 75\% accuracy.

We first considered a population in which all neurons have similar stimulus tuning and hence fire preferentially to the same stimulus (this is often referred to as positive stimulus correlations \cite{gawne}).  As above, we varied the triplet interaction parameters of a third-order maximum entropy model (Equation ~\eqref{e.triplet}), re-tuning the lower-order interaction parameters each time to keep constant the population's mean activity and pairwise correlations. Specifically, triplet interaction parameters were increased or decreased to explore each of the four quadrants in Figures 2 and 3. 

\begin{figure}[!t]
\begin{centering}
\leavevmode
\includegraphics[width=5in] {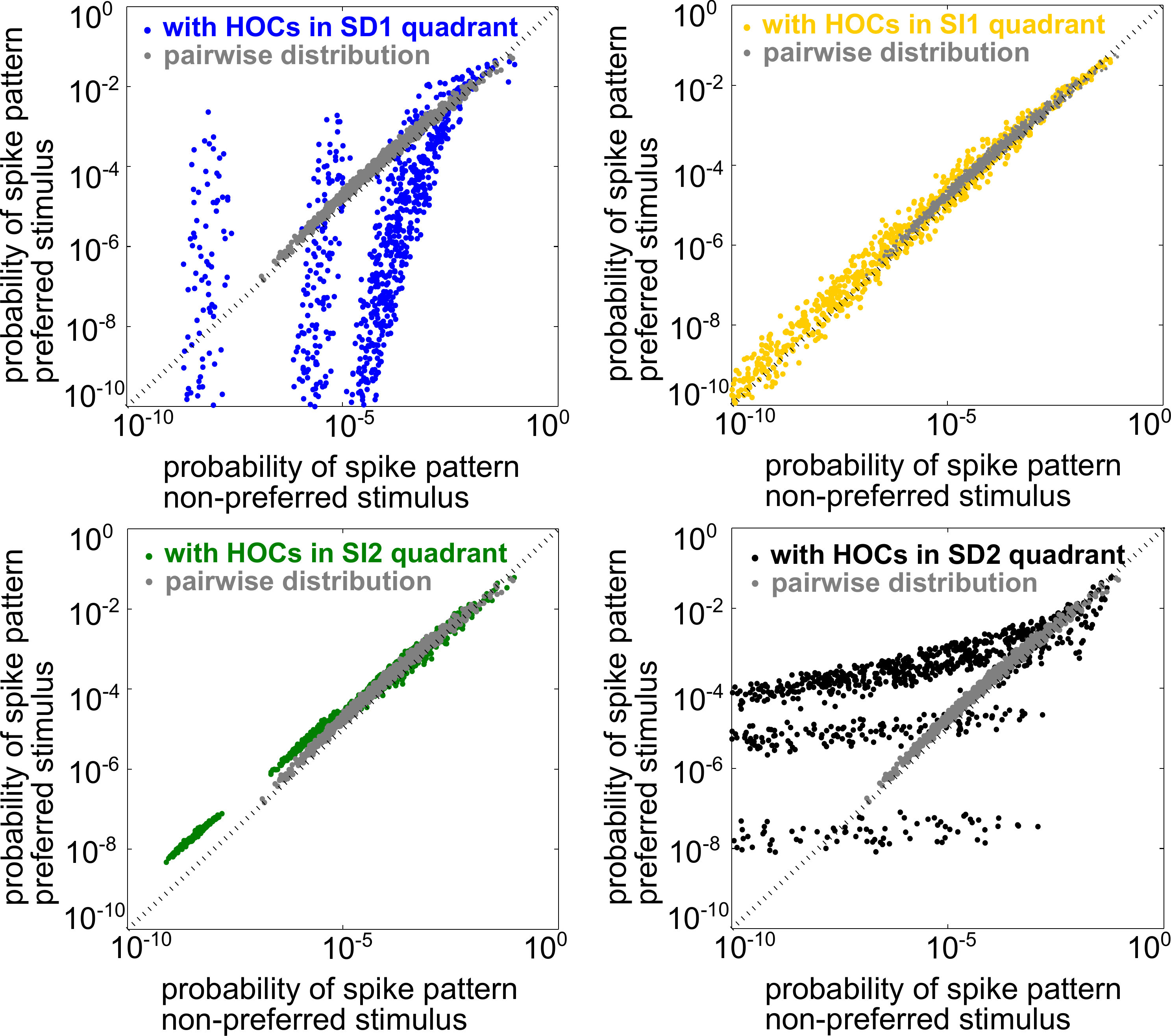}
\caption{Illustration of stimulus discriminability based on spike patterns in a heterogeneous neural population. Each point represents a different spiking pattern either for the pairwise model (grey, same model for all panels) or one with triplet correlations from one of the four quadrants in Figure 3A. The firing rates and pairwise correlations are identical for all five populalations. The axes represent the probability of that spiking pattern under each stimulus. The triplet statistics drawn from quadrants SD1 and SD2 lead to better stimulus discrimination, since the points lie far from the identity line (see text).}
\vspace{ 0 in}
\hspace{-.4 in}
\end{centering}
\end{figure}

Since the spiking statistics are heterogeneous across the population, mutual information must be computed using the response distributions over all spiking patterns rather than simply over spike counts, as in the homogeneous case. In this setting, the two stimuli are the most discriminable when the population spike patterns have the most different frequencies under each stimulus. To illustrate this, Figure 4 shows scatter plots of the probability of every firing pattern under the preferred versus the non-preferred stimulus.  Good discriminability between the stimuli therefore corresponds to points lying far from the identity line.  The figure shows probabilities for four example populations, each having the same lower-order statistics but differing in triplet interaction terms.  The four populations correspond to the four stimulus-dependent (SD) and stimulus-independent (SI) cases introduced for homogeneous populations above.  For comparison, grey points show responses for the pairwise model. The presence of triplet correlations changed spike pattern probabilities in each case.  However, these changes only significantly improved discriminability when they are stimulus-dependent.  Stimulus-independent triplet correlations failed to significantly affect discrimination because they change the probabilities in a similar way for each stimulus.  In sum, it appears that the same rule of thumb that we found for the homogeneous populations also applies here:  stimulus-dependent triplet correlations can significantly improve population coding in cases where stimulus-independent correlations will have little effect.

\begin{figure}[p!]
\begin{centering}
\leavevmode
\includegraphics[width=4.7in] {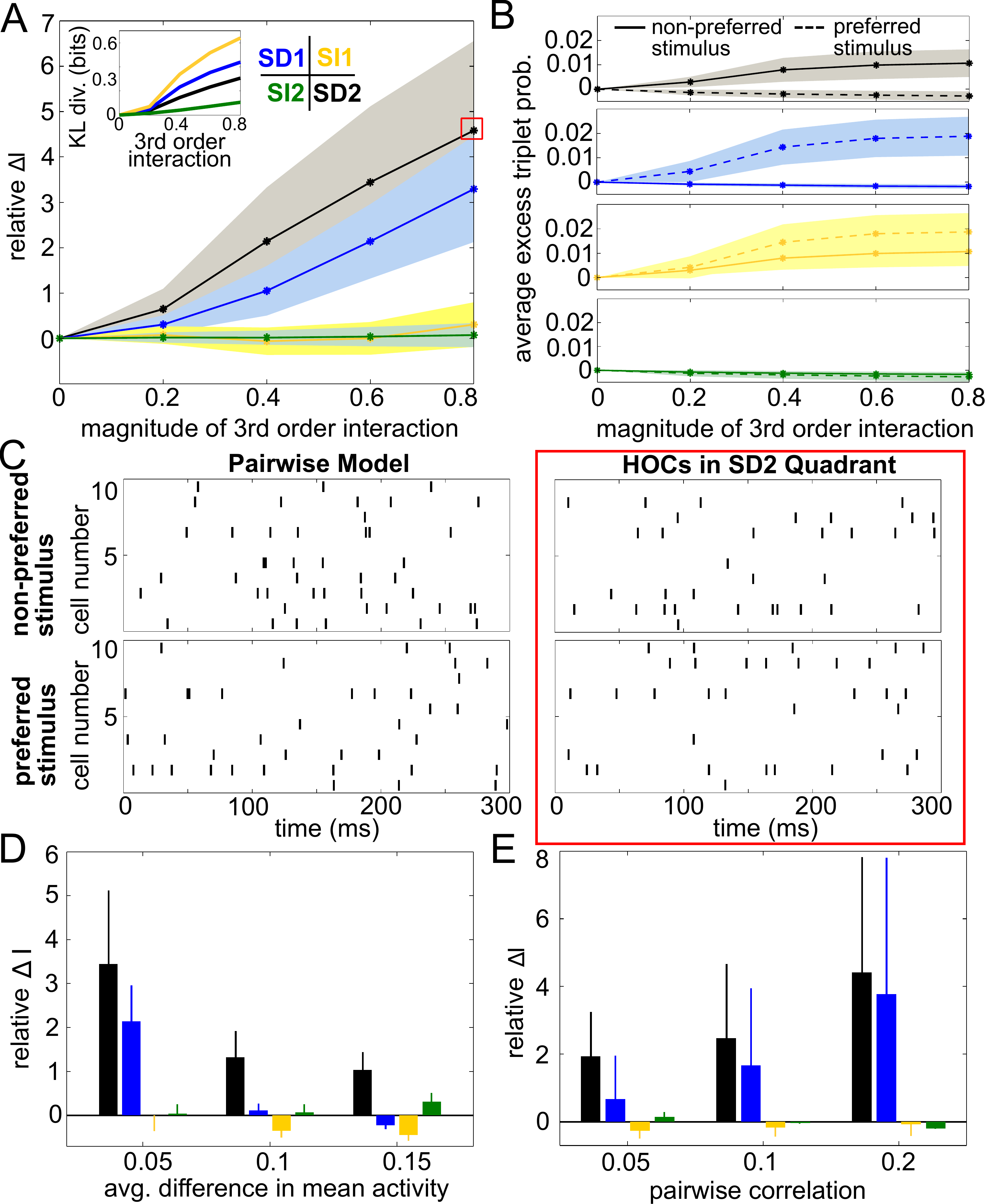}
\caption{Impact of triplet correlations on stimulus coding for populations with heterogeneous spiking statistics and similar stimulus tuning for all cells. (A) Relative increase in information $\Delta I$, averaged across 24 populations with different single-cell and pairwise statistics.  $\Delta I$ is plotted against the magnitude of the third order interactions $G_{ijk}^{(m)}$, as the magnitudes of these interactions increase within the four different quadrants (see text).  Colors correspond to the quadrants indicated in Figure 3A.  Average discrimination accuracy over the 24 pairwise models is 60\%. The average correlation coefficient is 0.05 and the average difference between the probability of a spike under each stimulus is 0.05. The inset shows the average Kullback-Leibler divergence in bits between the triplet models and their pairwise maximum entropy fits. (B) Excess triplet probability for the non-preferred (solid lines) and preferred (dashed lines) stimuli, averaged over all triplets. (C) Raster plots for the population marked with a red box in A, and the pairwise model. Note that the triplet correlations do not create large population-wide events immediately apparent by eye. (D) Relative increase in information over varying $\Delta\mu$ with average correlation of $\rho = 0.05$. The average baseline firing rate (to the non-preferred stimulus) was fixed to 0.05. (E) Relative increase in information as a function of average pairwise correlation. Here, the triplet interaction term is fixed to a magnitude of 0.6. Values are averaged over all firing rates (see Methods, Heterogeneous Populations). All error bars and shading represent standard deviation.}
\vspace{ 0 in}
\hspace{-.4 in}
\end{centering}
\end{figure}

To test this idea, we next computed the coding effect of triplet correlations in population models with a range of spiking statistics.  Figure 5A shows the relative increase in encoded information compared to the pairwise maximum entropy models (Equation~\eqref{e.deltaI}). Because of our focus on small populations, we are able to calculate mutual information exactly without need for entropy estimators. Results were averaged over 24 random draws of firing rates and pairwise correlation matrices (see Methods, Heterogeneous Populations). Stimulus-dependent triplet correlations produced a significant effect while stimulus-independent triplet correlations did not, and again the optimal strategy that we found was to increase triplet spiking for the non-preferred stimulus and decrease triplet spiking for the preferred stimulus (region SD2).  Figure 5B verifies that the triplet interaction term ($G^{(m)}$ in Equation ~\eqref{e.triplet}) has the expected effect on the averaged excess triplet spike probability  ($\kappa_{ijk}$, from Equation ~\eqref{e.kappa}).  

Example rasters from a population in region SD2 (red box in Figure 5A) and the corresponding pairwise model are shown in Figure 5C. Despite the fivefold increase in mutual information, the effect of the added triplet correlations on spike rasters appears subtle to the eye. The similarity of the pairwise firing pattern distributions and the triplet distributions can be measured by the Kullback-Leibler (KL) divergence, which calculates the average difference between the log-likelihood of each firing pattern under the triplet and pairwise distributions. A large KL divergence indicates that the pairwise model would fit the neural data poorly if the triplet model were the ``true" distribution of firing patterns. The inset in Figure 5A shows that even a population with a fourfold increase in mutual information has a relatively low KL divergence of only 0.2, which is approximately the KL divergence between the experimental recordings and pairwise fit in salamander retina in \cite{ganmor}. Note that large KL divergence does not not necessarily correlate with a large increase in information. For example, populations in region SI1 have a KL divergence of up to 0.4 but minimal effect on discrimination. This fact is also illustrated in Figure 4: triplet correlations modify the firing pattern probabilities (yellow points) so that they are very different from the pairwise models (gray points), but they lie distributed around the identity line, showing that the firing pattern probabilities are similar between stimuli.

Over a variety of parameter choices, stimulus-dependent triplet statistics continued to have a strong effect on information. Figure 5D shows the relative increase in information as the difference between the stimulus-conditioned firing rates increases, averaged over networks with different firing rates (see Methods, Heterogenous Populations). The effect of triplet correlations decreased as the stimulus-conditioned means become more different because the response distributions are less overlapping; however, region SD2 continued to strongly enhance correlations while other regions have smaller effect. Finally, panel E shows the relative increase in information for networks with increasing pairwise correlations. In highly correlated networks, any stimulus-dependent triplet correlations (both region SD1 and region SD2) strongly increased information. 


\begin{figure}[t]
\begin{centering}
\leavevmode
\includegraphics[width=6.5in] {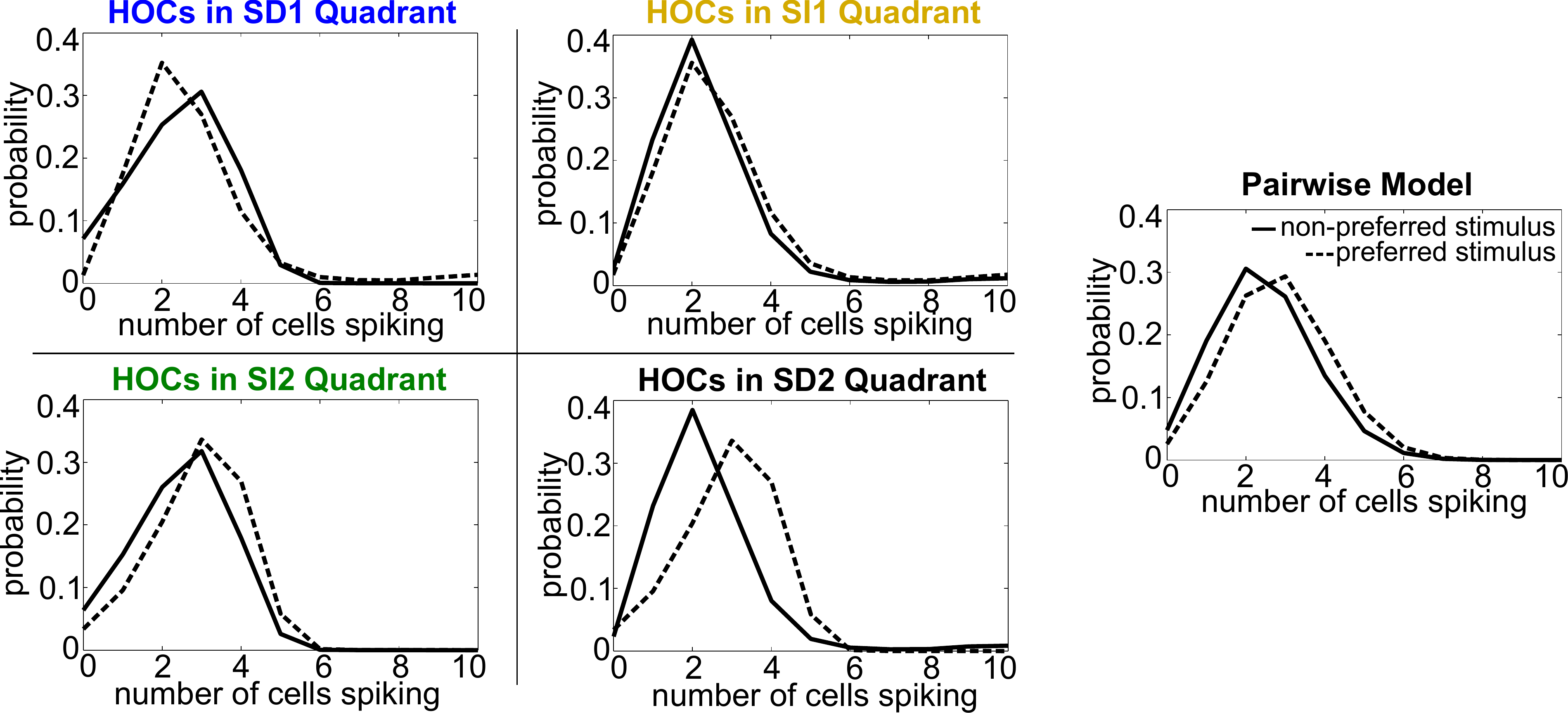}
\caption{Spike count histograms for five sample populations, all of which share the same heterogeneous lower-order statistics. Panels show the pairwise model (in which $G^{(m)}=0$) and the four different quadrants of triplet interactions ($G^{(m)}=\pm0.8$). Parameters are taken from the red box in Figure 5 but are reduced from probabilities of spiking patterns to distributions of spike counts. The average pairwise correlation coefficient is $\rho = 0.05$ and the average difference between the probability of a spike under each stimulus is $\Delta \mu = 0.05$.}
\vspace{ 0 in}
\hspace{-.4 in}
\end{centering}
\end{figure}

Intuitively, these effects follow the predictions from the schematic in Figure 2 that stimulus-dependent triplet correlations enhance discrimination by skewing the response distributions. Illustrating this, Figure 6 shows a reduction of the distributions to the population spike count distributions for the four quadrants in one population from Figure 5A. The spike count response distributions are skewed away from each other in region SD2 (and to a lesser extent in SD1), whereas stimulus-independent statistics (in SI1 or SI2) shape the distributions in the same direction. Even though the intuition in Figure 2 describes the effects of skewing distributions of the population spike \emph{count}, the findings here agree with the trends shown for stimulus information based on spike \emph{patterns}. In fact, Figure 7A shows the strong correlation between the raw increase in mutual information in the individual firing patterns (abscissa) and in the population-wide spike count (ordinate) for all populations in Figure 5A. This correlation is only guaranteed when the triplet correlations are all within the same quadrant (defined in Figure 2) and is not generally true for randomly generated population statistics (Figure 7B).

\begin{figure}[!t]
\begin{centering}
\leavevmode
\includegraphics[width=5.5in] {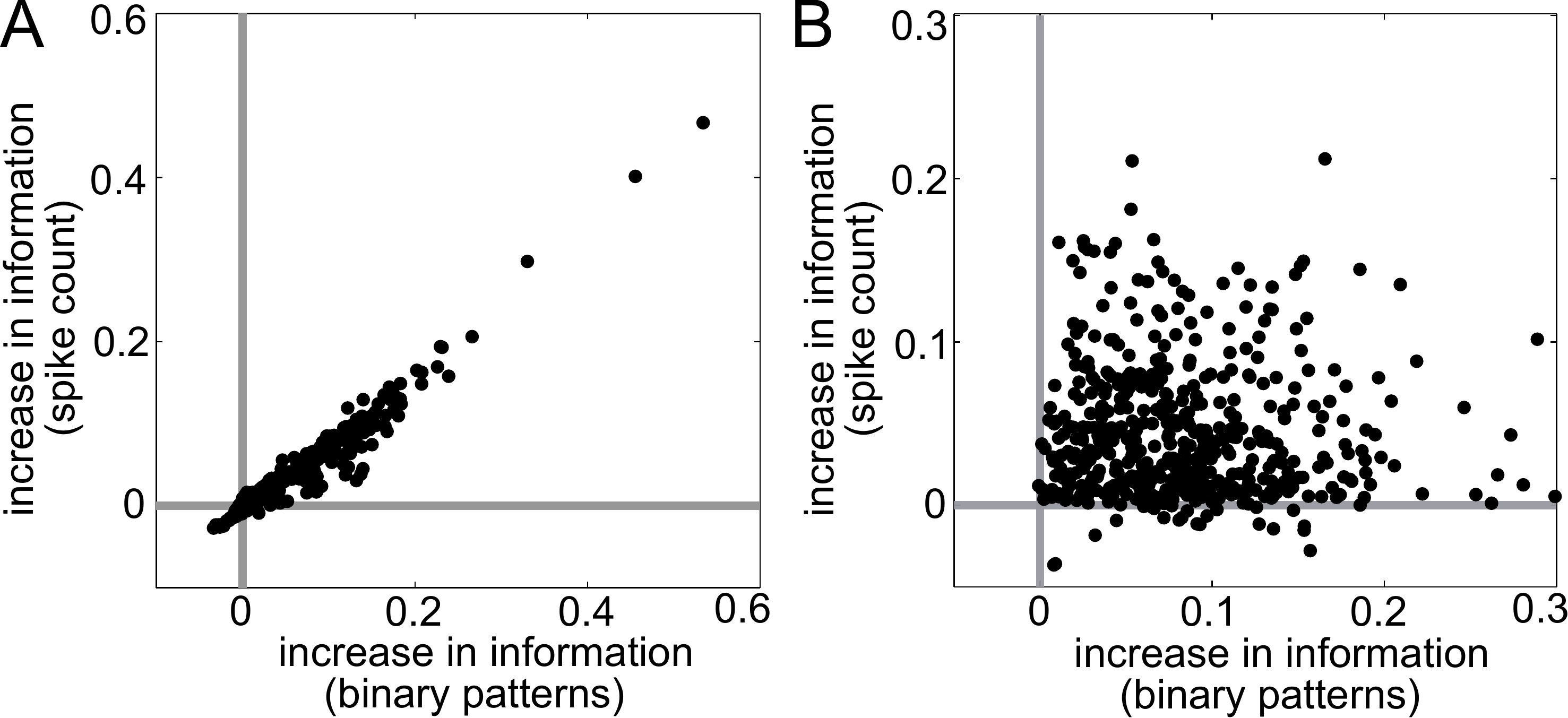}
\caption{Raw increase in information relative to pairwise ($I-I_{PW}$) for (A)  all populations shown in Figure 5A and (B) 500 populations with random interaction parameters. Abscissa represents the increase in information over all firing patterns, while the ordinate shows the increase in information in the spike count distributions.}
\vspace{ 0 in}
\hspace{-.4 in}
\end{centering}
\end{figure}

\begin{figure}[!p]
\begin{centering}
\leavevmode
\includegraphics[width=4.9in] {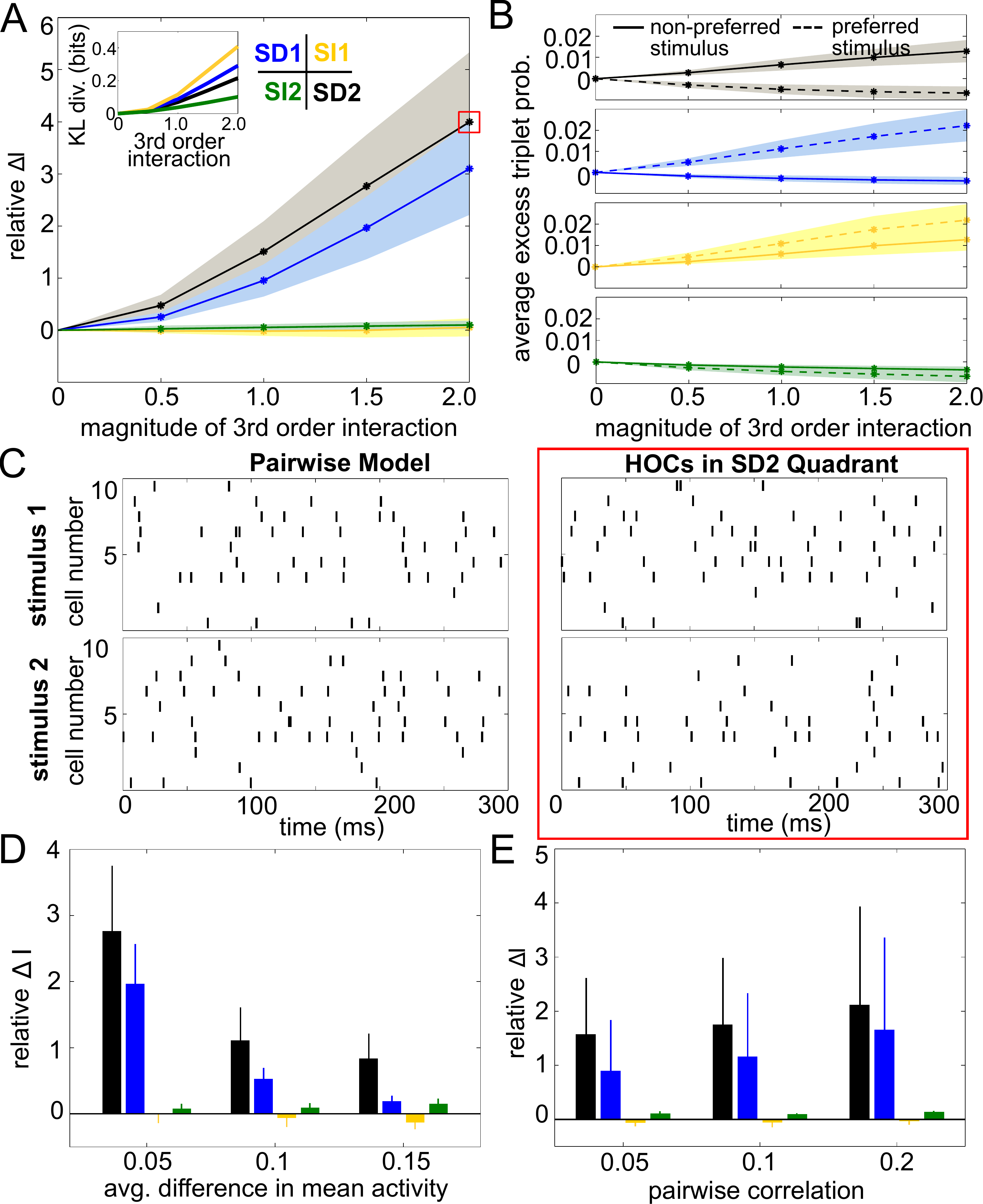}
\caption{Impact of triplet correlations on stimulus coding for populations with heterogeneous spiking statistics and different stimulus tuning for subgroups of cells.  (A)  Relative increase in information $\Delta I$, averaged across 24 populations with different single-cell and pairwise statistics.  $\Delta I$ is plotted against the magnitude of the third order interactions $G_{ijk}^{(m)}$, as the magnitudes of these interactions increase within the four different quadrants (see text). Note that stronger triplet interaction terms than in Figure 5A are required to have an effect on information because fewer triplets are varied in this case. Colors correspond to the quadrants indicated in Figure 3A. Average discrimination accuracy over the 24 pairwise models is 60\%. The average correlation coefficient is 0.05 and the average difference between the probability of a spike under each stimulus is 0.05. The inset shows the average Kullback-Leibler divergence in bits between the triplet models and their pairwise maximum entropy fits. (B) Excess triplet probability for the non-preferred (solid lines) and preferred (dashed lines) stimuli, averaged over all triplets. (C) Raster plots for the population marked with a red box in A, and the pairwise model. Note that the triplet correlations do not create large population-wide events immediately apparent by eye. (D) Relative increase in information over varying $\Delta\mu$ with average correlation of $\rho = 0.05$. The average baseline firing rate (to the non-preferred stimulus) was fixed to 0.05. (E) Relative increase in information as a function of average pairwise correlation. Here, the triplet interaction term is fixed to a magnitude of 1.5. Values are averaged over all firing rates (see Methods, Heterogeneous Populations). All error bars and shading represent standard deviation.}
\vspace{ 0 in}
\hspace{-.4 in}
\end{centering}
\end{figure}

Finally, we tested whether the same effects of triplet correlations on stimulus information would occur in populations with more diverse stimulus tuning. Towards this end, we split the populations into two groups of cells, each preferring a different stimulus. 
Within each subgroup, all triplets had the same interaction parameter $G_{ijk}^{(m)}$. The magnitude of this triplet interaction term was varied while the sign was fixed in accordance with the four quadrants in Figure 3A. For example, in region SD1, $G_{ijk}^{(m)}$ for a particular triplet is positive under the preferred stimulus for neurons $i$, $j$, and $k$, and is negative under the non-preferred stimulus for those neurons. The triplet interaction terms for triplets composed of cells drawn from both subgroups were set to zero. That is, nonzero triplet interactions only occurred for cells with similar stimulus tuning, a choice consistent with empirical observations of triplet correlations being localized to nearby cortical minicolumns \cite{ohiorhenuan}. We also tried manipulating all triplets regardless of subgroup, and saw a similar increase in information for stimulus-dependent triplet correlations, but the scale of the effect was significantly smaller (data not shown). 

\begin{figure}[!t]
\begin{centering}
\leavevmode
\includegraphics[width=6in] {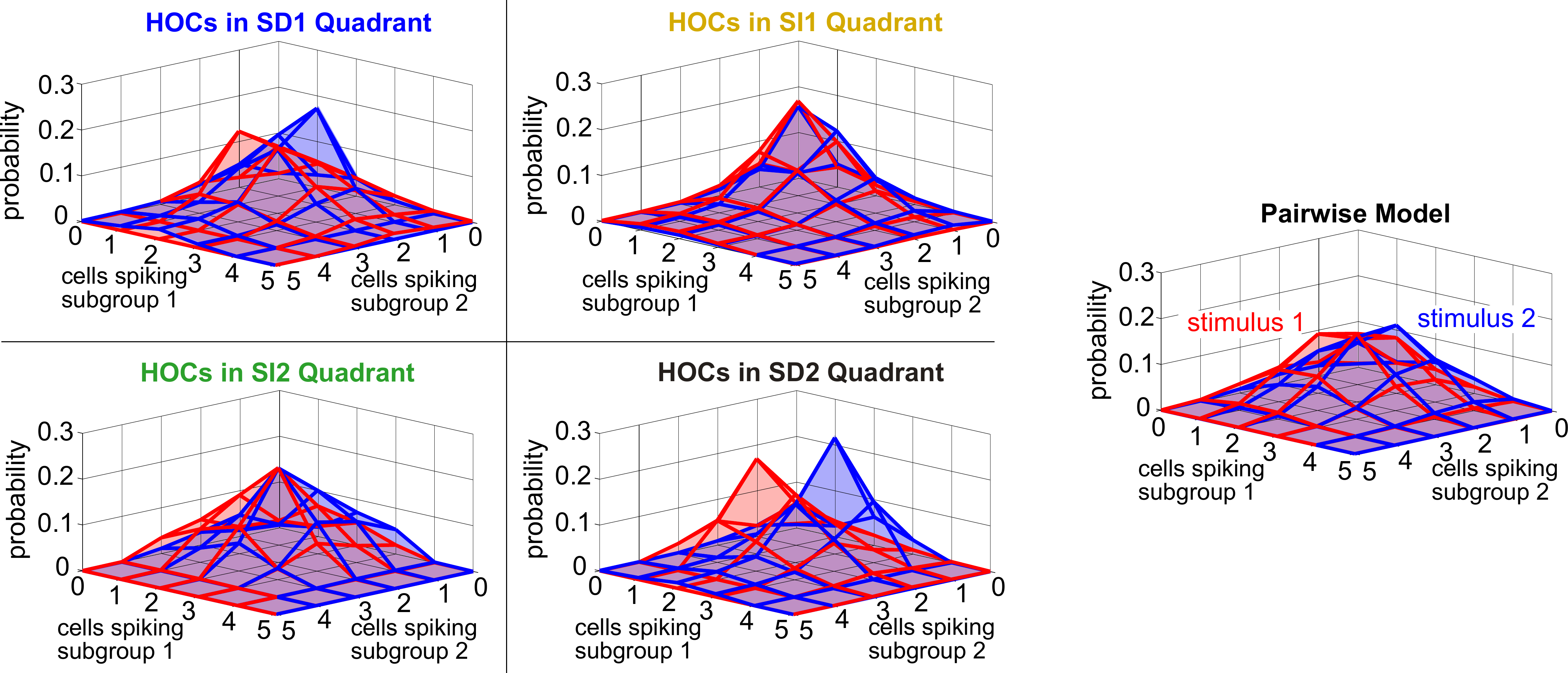}
\caption{Spike count histograms for five sample populations with dissimilar stimulus tuning, all of which share the same inhomogenous lower-order statistics. Panels show the pairwise model, right, (in which $G^{(m)}=0$) and the four different quadrants of triplet interactions, left, ($G^{(m)}=\pm2.0$). Parameters are taken from the red box in Figure 8 but are reduced from probabilities of spiking patterns to spike counts. In particular, the average pairwise correlation coefficient is 0.05 and the average difference between the probability of a spike under each stimulus is 0.05.}
\vspace{ 0 in}
\hspace{-.4 in}
\end{centering}
\end{figure}

Results were qualitatively the same as before (Figure 8AD). Stimulus-independent triplet correlations made little difference on the discriminability of the stimuli.  Meanwhile, the largest increase in information occurred in region SD2, when the frequency of triplet spikes within each subgroup was depressed under the preferred stimulus and enhanced under the non-preferred stimulus. The changes in triplet spiking from case to case continued to have only a subtle impact on the raster plots (Figure 8C). Finally, stimulus-dependent correlations in region SD2 continued to have a strong effect on networks with different stimulus-conditioned firing rates and different average pairwise correlations (Figure 8DE).

Finally, we asked whether the same intuition that we have developed throughout this paper, about how triplet correlations impact stimulus encoding by skewing distributions of population spike counts, also applies here. Because the two subgroups differ in stimulus selectivity, we did not group their spikes into a single count; instead, we considered the spike counts of the two subgroups separately. The resulting two-subgroup spike count histograms are shown in Figure 9.  These provide insight into how the triplet correlations shape the response distributions.  The triplet correlations in region SD2 skew the two-dimensional response distributions away from each other, 
allowing the stimuli to be better distinguished. 
Stimulus-independent triplet correlations, however, again shape the distributions in the same way for both stimuli.  We conclude that, even for our inhomogenous populations with diverse stimulus tuning, the intuition developed in Figure 2 describes how triplet correlations can affect the encoding of preferred versus non-preferred stimuli.

\subsection*{How much data is necessary to estimate HOCs?}

\begin{figure}[!t]
\begin{centering}
\leavevmode
\includegraphics[width=4.8in] {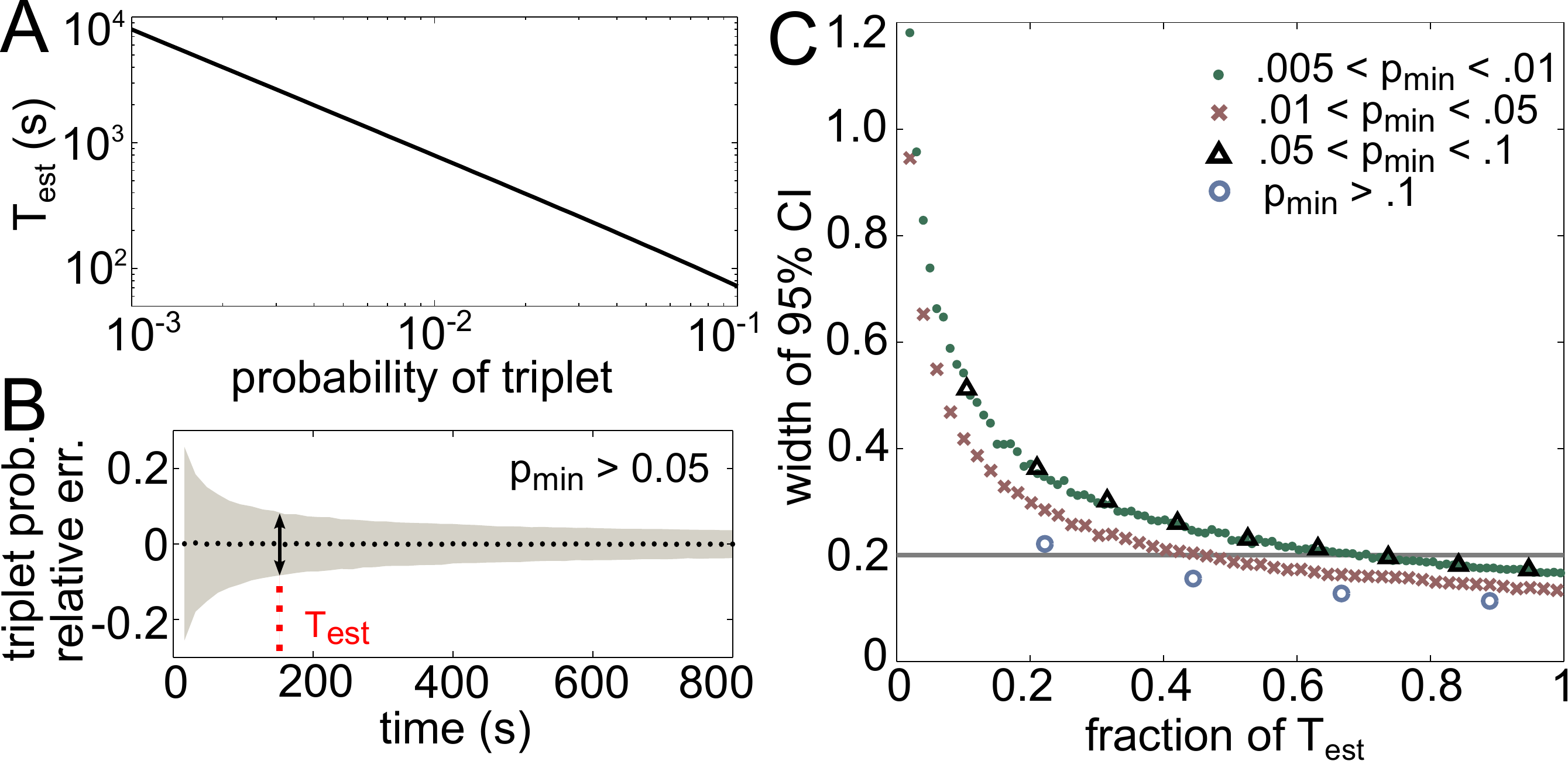}
\caption{Amount of data necessary to accurately estimate triplet frequencies. In all panels, $\alpha = 0.1$ (this represents the prescribed relative error; see text). (A) $T_\text{est}$ scales as a power law for small probabilities; here it is plotted in seconds, assuming 20 ms time bins. (B) Triplet probabilities were estimated from samples of 1000 triplet maximum entropy models with randomly chosen interaction parameters using different amounts of simulated data. Black dotted line shows the average relative error for all triplets with frequency greater than $p_\text{min}=0.05$ as a function of time used in the estimation (assuming 20 ms time bins). Grey funnel represents the 95\% confidence interval. Dotted red line shows $T_\text{est}$ calculated from Equation ~\eqref{e.test}. As expected, the width of the confidence interval here (denoted by the arrows) is 2$\alpha=0.2$. (C) Width of 95\% confidence interval (CI) plotted as a fraction of $T_\text{est}$ for four choices of $p_\text{min}$. All widths are below $2\alpha$ by time $T=T_\text{est}$.}
\vspace{ 0 in}
\hspace{-.4 in}
\end{centering}
\end{figure}

Above, we have seen when and how triplet spiking statistics can have a significant impact on discrimination in neural populations. To characterize the effect of higher-order correlations in data, accurate measurements of the frequencies of spiking patterns are crucial. An essential source of difficulty in observing HOCs is the amount of data required. Since synchronous spiking events are relatively infrequent, they require longer recordings or many trials to measure. We estimated the amount of data that is required to measure the likelihood of a triplet of neurons spiking synchronously within a relative error of $\alpha$ by bounding the 95\% confidence interval of any triplet of probability larger than $p_\text{min}$ (see Methods for details). This gives the following equation:
\begin{equation}
T_\text{est} = \frac{1-p_\text{min}}{p_\text{min}\left(\frac{\alpha}{2}\right)^2}.
\end{equation}
$T_\text{est}$ provides a lower bound on the number of binned activity patterns that are necessary to measure all triplets with frequencies of $p_\text{min}$ or greater within a relative error of $\alpha$. The choice of bin size is an important issue that we do not address here, as it does not affect these results. Figure 10A illustrates the dependence of $T_\text{est}$ on $p_\text{min}$ for a relative error of $10\%$, or $\alpha=0.1$ (plotted in seconds assuming time bins of 20 ms). Note the logarithmic scaling on the axes: for example, only 220 seconds of data would be necessary to estimate the average triplet probabilities in Figure 8C (right panels), but over two hours are needed to estimate the least frequent triplets.

To test the tightness of the bound, we generated third order maximum entropy distributions with random interaction parameters and calculated the probability of three neurons firing synchronously from independent samples from the distribution. Figure 10B shows the mean relative error (black dots) and two standard errors of the mean (gray funnel) for all triplets with sample probability greater than $p_\text{min} = 0.05$. At the estimate $T_\text{est}$, the width of the 95\% confidence interval is around $2\alpha$, as predicted. The estimate is shown to be accurate for several ranges of $p$ in Figure 10C; in fact, the estimate is conservative, because probabilities larger than $p_\text{min}$ will require even less data. This formula can be helpful for designing experiments to detect infrequent spiking events; or alternately, given a data set, this formula specifies which spiking patterns have sample frequencies that are large enough to be relatively accurately determined.

\section{Discussion}

The spiking patterns that neural populations produce in response to a given stimulus are variable, and this variability is correlated from cell to cell.   There has been extensive work on how these correlations impact the fidelity with which a population encodes its stimuli, but most of this work has focused on correlations between pairs of cells.  Here, we held such pairwise correlations fixed and explored the impact of triplet correlations, which have recently been observed in multiple brain areas, on discriminating between preferred versus non-preferred stimuli in small populations of neurons.  

Starting with homogeneous populations and working through those with progressively more diverse properties, we found that a common set of principles governed the impact of triplet correlations on the discrimination of stimuli.  
When triplet spike correlations were either increased or decreased relative to the level occurring in a null ``pairwise model," and this increase or decrease occurred similarly for both stimuli, there was little impact on coding accuracy. However, stimulus-dependent triplet correlations significantly enhanced coding, by shaping the response distributions to reduce their overlap. In particular, when pairwise correlations were low, the greatest improvements were found when triplet spike correlations were decreased for the preferred stimulus, and increased for the non-preferred stimulus. Intuitively, these effects can be understood as skewing the stimulus-conditioned spike count distributions away from or towards each other (as in Figure 2). We showed that this intuition is fruitful even when considering the information encoded in spiking patterns of heterogenous populations with more diverse tuning properties. 

Thus, if triplet correlations vary with stimuli, models that only take pairwise statistics into account could significantly underestimate the information represented in neural populations, at least in the cases we study here. Importantly, the presence of triplet correlations can be easily overlooked despite their potentially large impact on stimulus encoding: for example, some measures of coding accuracy, such as the optimal linear estimator, do not incorporate higher-order correlations. Second, higher-order spiking statistics are difficult to observe from raster plots alone (as in Figure 1). Finally, even direct measurements may be impractical in some cases as long recordings are necessary to reliably sample infrequent spiking events. With an eye toward future experiments, we provide an estimate in Equation~\eqref{e.test} of how much data is required to accurately measure higher-order statistics within a given relative error.

Whether neural circuits actually exploit our finding that stimulus-dependent triplet correlations can strongly improve coding remains unknown. At the level of pairs of cells, correlations in cortex are modulated by task relevance \cite{jeanne} and attention \cite{cohen09}; beyond-pairwise interactions can be modulated during motion preparation in motor cortex of awake macaques \cite{shimazaki}. On the other hand, in \cite{ohiorhenuan}, higher-order spiking correlations in anesthetized macaque visual cortex were found to be negative regardless of stimulus (as in region SI2 in Figure 3A).  In agreement with our general theory, these triplet correlations had no measurable effect on encoded information. 

A natural question that arises from our findings is the mechanistic origin of stimulus-dependent higher-order correlations.  While common input is a prime candidate for the generation of HOCs in general, stimulus-dependence might stem from intrinsic nonlinearities such as thresholding or spike generation \cite{macke, barreiro, zylberberg}. On the other hand, if triplet correlations act similarly under differing stimuli, they may have no impact on coding; intriguingly, however, they may serve a complimentary purpose such as sparsifying the neural code \cite{ohiorhenuan}. Moving forward, one could test experimentally how higher-order correlations are modulated during learning in animals that are trained to discriminate between similar stimuli. If the population spiking statistics adapt so that triplet correlations are strongly stimulus-dependent after training, this would be an indicator that neural systems can use higher-order correlations to their advantage to better discriminate between similar stimuli.

Our study had a number of simplifications and limitations that will be addressed in future work.  First, we chose to study discrimination between pairs of stimuli, but the approach could be extended to encoding of multiple stimuli. Second, because we were interested in isolating the effect of triplet correlations, we held pairwise statistics constant from one stimulus to the next.  Our intuition may generalize, however, to cases where these pairwise correlations also change with stimuli.  In the schematic of Figure 2, increasing correlations between pairs of neurons will change the variances of the population spike count, but will not change the effect of oppositely-skewing the spike count distributions once the lower-order moments are fixed. 
However, it would be interesting to study varying pairwise and higher-order statistics together. 

Furthermore, because maximum entropy models assume that responses are stationary in time, they are generally used to characterize zero-lag correlations rather than more complicated temporal dependencies. While the models can in theory be extended to include spatiotemporal patterns \cite{marre}, the added dimensionality is a major hurdle to overcome.   

This leads to perhaps the strongest limitation of our study --- we study only relatively small population sizes.  This is due in part to the computational expense of tuning maximum entropy models with order $N^2$ parameters, while varying triplet interaction terms systematically. Exact calculations of mutual information also become intractable in large populations, as the probabilities of $2^N$ states must be enumerated.  For certain sensory coding problems, population sizes close to  the $N=10$ we used may be the relevant order of magnitude. For instance, only eight directionally selective ganglion cells encode motion at each retinal location \cite{amthor}. In other applications, this number is clearly insufficient. 

A comprehensive analysis of the impact of higher order correlations on coding in larger cell populations will be the topic of future research.  We expect the intuition we developed based on the skewness of response distributions to hold for larger populations, as long as the triplet interaction parameters are restricted to fall squarely in one of the four quadrants in Figure 3A (and are therefore relatively homogenous across the population).   We have confirmed that, for fixed triplet statistics (excess triplet spiking $\kappa$) the relative increase in information due to triplet correlations can remain stable as $N$ increases, at least for homogenous populations (supplementary figure S1).  However, for the setting of this paper --- in which we fix pairwise correlations and firing rates to relatively low values and assume that triplet correlations exist among every triplet within the population --- the range of possible triplet correlations is likely to decrease with $N$, and this may limit their possible impact on encoded information.  Thus, the present work is best thought of as investigating the impact of triplet correlations in small subpopulations sharing similar tuning preferences, perhaps as for the localized triplet correlations found in primate cortex
~\cite{ohiorhenuan}.

To fully understand encoding in neural circuits, it is essential to characterize the functional interactions between different groups of neurons, and how they change with external stimuli. With this work, we have made a first step toward extending this program to incorporate beyond-pairwise spike correlations.  With ongoing advances in high-density recordings and large-scale data analysis, we can look forward to an increasingly unified theory of how neural covariability at all orders impacts coding.

\section{Acknowledgements}
We thank Kreso Josic, Michael Buice, Ali Weber, and Braden Brinkman for helpful comments on the manuscript.

\bibliographystyle{plos2009}
\bibliography{mybib}

\newpage

\begin{figure}[!p]
\begin{centering}
\leavevmode
\includegraphics[width=7in] {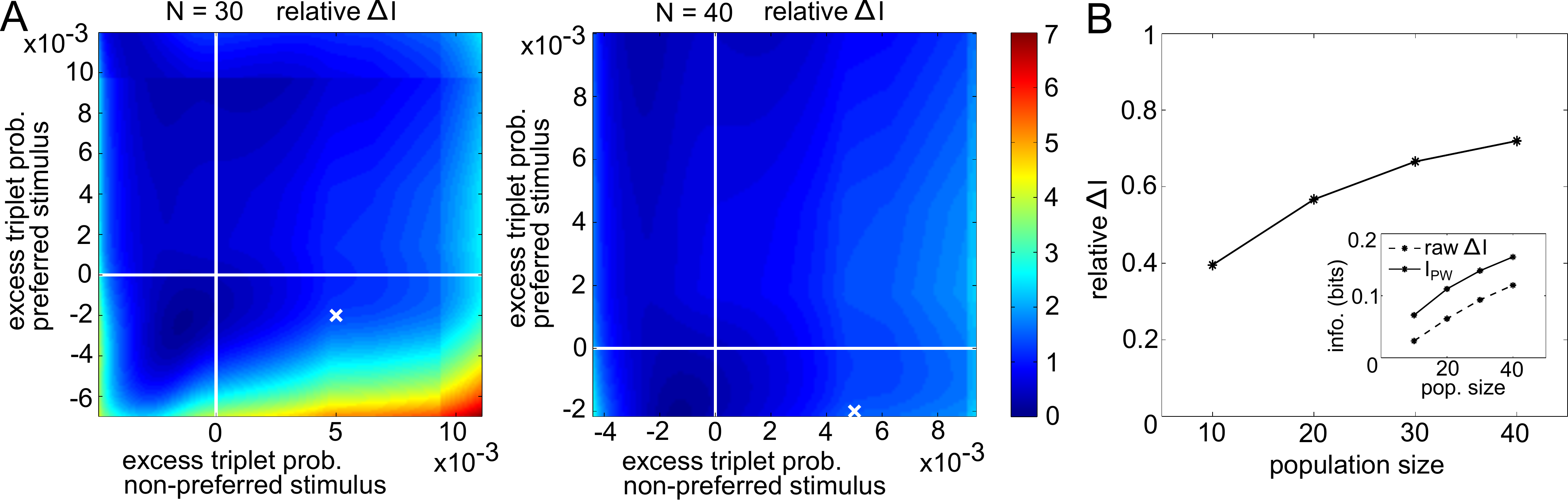}
\caption{Supplementary figure S1. Effect of population size on mutual information in homogenous populations. (A) Relative increase in information as the excess triplet probability is varied for responses to the preferred and non-preferred stimuli, shown for populations of $N=30$ (left) and $N=40$ (right) cells. Note that the range of triplet correlations is smaller for the larger population. This is because, as $N$ increases, there are tighter constraints on the possible values of triplet correlations that can be attained homogeneously across every triplet in the population, while still maintaining the same (low) predefined firing rates and pairwise correlations.
Still, in the region of overlap, the strength of the impact on mutual information is similar in magnitude in both plots. Here, firing rates and pairwise correlations are fixed to: $\mu_1 = 0.25$, $\mu_2 = 0.35$, $\rho = 0.05$. Compare with the plot of raw mutual information (as opposed to the relative increase in information) in 10-cell populations that is shown in Figure 3A. (B) Relative increase in information (black curve) for fixed triplet correlations and lower-order statistics, for increasing population size. Specifically, the values of the triplet correlations were: $\kappa=.005$ for the non-preferred stimulus, and $\kappa=-.002$ for the preferred stimulus, corresponding to the ``x" in panel A. The impact of the triplet statistics on mutual information grows with population size.  Inset shows $I_{PW}$, the mutual information between the pairwise distribution and the stimuli (solid line) and $\Delta I$, the raw increase in information due to triplet correlations (dashed line) for varying population size.}
\vspace{ 0 in}
\hspace{-.4 in}
\end{centering}
\end{figure}

\end{document}